\documentclass[journal]{IEEEtran}

\usepackage{cite}
\usepackage{amsmath,amssymb,amsfonts}
\usepackage{algorithmic}
\usepackage{graphicx}
\usepackage{textcomp}
\usepackage{xcolor}
\usepackage{multirow}
\usepackage{float}
\usepackage{array}
\newcolumntype{P}[1]{>{\centering\arraybackslash}p{#1}}
\newcolumntype{M}[1]{>{\centering\arraybackslash}m{#1}}
\usepackage{soul}

\usepackage[font=scriptsize]{subcaption}
\usepackage[font=footnotesize]{caption}
\usepackage[acronyms,nonumberlist,nopostdot,nomain,nogroupskip]{glossaries}
\usepackage{hyperref}
\newcommand{\comment}[1]{}
\usepackage{booktabs}% http://ctan.org/pkg/booktabs

\usepackage{tikz}
\usepackage{pgfplots}
\pgfplotsset{compat=newest} 
\pgfplotsset{plot coordinates/math parser=false} 
\newlength\fheight
\newlength\fwidth
\usetikzlibrary{plotmarks,patterns,decorations.pathreplacing,backgrounds,calc,arrows,arrows.meta,spy,matrix}
\usepgfplotslibrary{patchplots,groupplots}
\usepackage{tikzscale}
\usepackage{hyperref}
\usepackage{siunitx}

\usepackage{multirow}
\usepackage{tikz-qtree}
\usetikzlibrary{trees} % this is to allow the fork right path
\usepackage{siunitx}
\usetikzlibrary{fadings}

\tikzfading[name=middle,
            top color=transparent!100,
            bottom color=transparent!100,
            middle color=transparent!20]

\usetikzlibrary{arrows,automata,calc,shapes, positioning,shadows,shadows.blur,shapes.geometric}

\newacronym{3gpp}{3GPP}{3rd Generation Partnership Project}
\newacronym{itu}{ITU}{International Telecommunication Union}
\newacronym{adc}{ADC}{Analog to Digital Converter}
\newacronym{5g}{5G}{5th generation}
\newacronym{6g}{6G}{6th generation}
\newacronym{aimd}{AIMD}{Additive Increase Multiplicative Decrease}
\newacronym{am}{AM}{Acknowledged Mode}
\newacronym{amc}{AMC}{Adaptive Modulation and Coding}
\newacronym{aqm}{AQM}{Active Queue Management}
\newacronym{awgn}{AGWN}{Additive White Gaussian Noise}
\newacronym{balia}{BALIA}{Balanced Link Adaptation}
\newacronym{bdp}{BDP}{Bandwidth-Delay Product}
\newacronym{bf}{BF}{beamforming}
\newacronym{cc}{CC}{Congestion Control}
\newacronym{cdf}{CDF}{Cumulative Distribution Function}
\newacronym{cn}{CN}{Core Network}
\newacronym{cqi}{CQI}{Channel Quality Information}
\newacronym{cp}{CP}{Control Plane}
\newacronym{csirs}{CSI-RS}{Channel State Information - Reference Signal}
\newacronym{dc}{DC}{Dual Connectivity}
\newacronym{rb}{RB}{Resource Block}
\newacronym{dce}{DCE}{Direct Code Execution}
\newacronym{dci}{DCI}{Downlink Control Information}
\newacronym{udp}{UDP}{User Datagram Protocol}
\newacronym{dl}{DL}{Downlink}
\newacronym{dmr}{DMR}{Deadline Miss Ratio}
\newacronym{dmrs}{DMRS}{DeModulation Reference Signal}
\newacronym{e2e}{E2E}{End-to-End}
\newacronym{ppp}{PPP}{Poission Point Process}
\newacronym{si}{SI}{Study Item}
\newacronym{ecn}{ECN}{Explicit Congestion Notification}
\newacronym{edf}{EDF}{Earliest Deadline First}
\newacronym{enb}{eNB}{eNodeB}
\newacronym{epc}{EPC}{Evolved Packet Core}
\newacronym{es}{ES}{Edge Server}
\newacronym{cav}{CAV}{Connected and Autonomous Vehicle}
\newacronym{fdma}{FDMA}{Frequency Division Multiple Access}
\newacronym{fdd}{FDD}{Frequency Division Duplexing}
\newacronym{upa}{UPA}{Uniform Planar Array}
\newacronym[firstplural=Radio Access Technologies (RATs)]{rat}{RAT}{Radio Access Technology}
\newacronym[firstplural=Radio Access Technology (RTs)]{rt}{RT}{Radio Technology}
\newacronym{fs}{FS}{Fast Switching}
\newacronym{isd}{ISD}{inter-site distance}
\newacronym{ftp}{FTP}{File Transfer Protocol}
\newacronym{gnb}{gNB}{Next Generation Node Base}
\newacronym{harq}{HARQ}{Hybrid Automatic Repeat reQuest}
\newacronym{hetnet}{HetNet}{Heterogeneous Network}
\newacronym{hh}{HH}{Hard Handover}
\newacronym{hol}{HOL}{Head-of-Line}
\newacronym{ia}{IA}{Initial Access}
\newacronym{imt}{IMT}{International Mobile Telecommunication}
\newacronym{iot}{IoT}{Internet of Things}
\newacronym{los}{LOS}{Line of Sight}
\newacronym{lte}{LTE}{Long Term Evolution}
\newacronym{m2m}{M2M}{Machine to Machine}
\newacronym{mac}{MAC}{Medium Access Control}
\newacronym{mc}{MC}{Multi-Connectivity}
\newacronym{mcs}{MCS}{Modulation and Coding Scheme}
\newacronym{mec}{MEC}{Mobile Edge Cloud}
\newacronym{mi}{MI}{Mutual Information}
\newacronym{mimo}{MIMO}{Multiple Input Multiple Output}
\newacronym{mmwave}{mmWave}{millimeter wave}
\newacronym{mptcp}{MPTCP}{Multipath TCP}
\newacronym{mr}{MR}{Maximum Rate}
\newacronym{mss}{MSS}{Maximum Segment Size}
\newacronym{mtd}{MTD}{Machine-Type Device}
\newacronym{mtu}{MTU}{Maximum Transmission Unit}
\newacronym{nfv}{NFV}{Network Function Virtualization}
\newacronym{vnf}{VNF}{ Virtualization Network Function}
\newacronym{sdn}{SDN}{Software Defined Networking}
\newacronym{nlos}{NLOS}{Non Line of Sight}
\newacronym{nlosb}{NLOSb}{Building Non Line of Sight}
\newacronym{nlosv}{NLOSv}{Vehicle Non Line of Sight}
\newacronym{nr}{NR}{New Radio}
\newacronym{ofdm}{OFDM}{Orthogonal Frequency Division Multiplexing}
\newacronym{pdcch}{PDCCH}{Physical Downlonk Control Channel}
\newacronym{pdcp}{PDCP}{Packet Data Convergence Protocol}
\newacronym{pdsch}{PDSCH}{Physical Downlink Shared Channel}
\newacronym{pdu}{PDU}{Packet Data Unit}
\newacronym{pf}{PF}{Proportional Fair}
\newacronym{pgw}{PGW}{Packet Gateway}
\newacronym{phy}{PHY}{Physical}
\newacronym{pbch}{PBCH}{Physical Broadcast Channel}
\newacronym[plural=\gls{mme}s,firstplural=Mobility Management Entities (MMEs)]{mme}{MME}{Mobility Management Entity}
\newacronym{prb}{PRB}{Physical Resource Block}
\newacronym{pss}{PSS}{Primary Synchronization Signal}
\newacronym{pucch}{PUCCH}{Physical Uplink Control Channel}
\newacronym{pusch}{PUSCH}{Physical Uplink Shared Channel}
\newacronym{rach}{RACH}{Random Access Channel}
\newacronym{ran}{RAN}{Radio Access Network}
\newacronym{red}{RED}{Random Early Detection}
\newacronym{rf}{RF}{Radio Frequency}
\newacronym{rlc}{RLC}{Radio Link Control}
\newacronym{rlf}{RLF}{Radio Link Failure}
\newacronym{rrc}{RRC}{Radio Resource Control}
\newacronym{rrm}{RRM}{Radio Resource Management}
\newacronym{rr}{RR}{Round Robin}
\newacronym{rs}{RS}{Remote Server}
\newacronym{rsrp}{RSRP}{Reference Signal Received Power}
\newacronym{rss}{RSS}{Received Signal Strength}
\newacronym{rtt}{RTT}{Round Trip Time}
\newacronym{rw}{RW}{Receive Window}
\newacronym{rx}{RX}{Receiver}
\newacronym{sa}{SA}{standalone}
\newacronym{sack}{SACK}{Selective Acknowledgment}
\newacronym{sap}{SAP}{Service Access Point}
\newacronym{sch}{SCH}{Secondary Cell Handover}
\newacronym{scoot}{SCOOT}{Split Cycle Offset Optimization Technique}
\newacronym{sdma}{SDMA}{Spatial Division Multiple Access}
\newacronym{sinr}{SINR}{Signal to Interference plus Noise Ratio}
\newacronym{sm}{SM}{Saturation Mode}
\newacronym{snr}{SNR}{Signal to Noise Ratio}
\newacronym{son}{SON}{Self-Organizing Network}
\newacronym{ss}{SS}{Synchronization Signal}
\newacronym{srs}{SRS}{Sounding Reference Signal}
\newacronym{sss}{SSS}{Secondary Synchronization Signal}
\newacronym{tb}{TB}{Transport Block}
\newacronym{tcp}{TCP}{Transmission Control Protocol}
\newacronym{tdd}{TDD}{Time Division Duplexing}
\newacronym{tdma}{TDMA}{Time Division Multiple Access}
\newacronym{tfl}{TfL}{Transport for London}
\newacronym{tm}{TM}{Transparent Mode}
\newacronym{prr}{PRR}{Packet Reception Ratio}
\newacronym{trp}{TRP}{Transmitter Receiver Pair}
\newacronym{tti}{TTI}{Transmission Time Interval}
\newacronym{ttt}{TTT}{Time-to-Trigger}
\newacronym{tx}{TX}{Transmitter}
\newacronym{ue}{UE}{User Equipment}
\newacronym{ul}{UL}{Uplink}
\newacronym{uml}{UML}{Unified Modeling Language}
\newacronym{um}{UM}{Unacknowledged Mode}
\newacronym{utc}{UTC}{Urban Traffic Control}
\newacronym{vm}{VM}{Virtual Machine}
\newacronym{rsrq}{RSRQ}{Reference Signal Received Quality}
\newacronym{rssi}{RSSI}{Received Signal Strength Indicator}
\newacronym{crs}{CRS}{Cell Reference Signal}
\newacronym{v2v}{V2V}{Vehicle-to-Vehicle}
\newacronym{v2i}{V2I}{Vehicle-to-Infrastructure}
\newacronym{v2n}{V2N}{Vehicle-to-Network}
\newacronym{v2x}{V2X}{Vehicle-to-Everything}
\newacronym{vn}{VN}{Vehicular Node}
\newacronym{dsrc}{DSRC}{Dedicated Short Range Communication}
\newacronym{ci}{CI}{context information}
\newacronym{voi}{VoI}{value of information}
\newacronym{gps}{GPS}{Global Positioning System}
\newacronym{qos}{QoS}{Quality of Service}
\newacronym{pqos}{PQoS}{Predictive Quality of Service}
\newacronym{qoe}{QoE}{Quality of Experience}
\newacronym{ml}{ML}{machine learning}
\newacronym{ai}{AI}{artificial intelligence}
\newacronym{ahp}{AHP}{Analytic Hierarchy Process}
\newacronym{lidar}{LIDAR}{Light Detection and Ranging}
\newacronym{sumo}{SUMO}{Simulation of Urban MObility}
\newacronym{wave}{WAVE}{Wireless Access in Vehicular Environment}
\newacronym{hd}{HD}{high definition}
\newacronym{c-its}{C-ITS}{Connected Intelligent Transportation System}
\newacronym{dash}{DASH}{Dynamic Adaptive Streaming over HTTP}
\newacronym{http}{HTTP}{HyperText Transfer Protocol}
\newacronym{nt}{NT}{non-terrestrial}
\newacronym{ntc}{NTC}{non-terrestrial communication}
\newacronym{ntn}{NTN}{non-terrestrial network}
\newacronym{haps}{HAPS}{High Altitude Platform Station}
\newacronym{hap}{HAP}{High Altitude Platform}
\newacronym{leo}{LEO}{Low Earth Orbit}
\newacronym{meo}{MEO}{Medium Earth Orbit}
\newacronym{geo}{GEO}{Geostationary Earth Orbit}
\newacronym{uav}{UAV}{Unmanned Aerial Vehicle}
\newacronym{nsat}{nSAT}{Nanosatellite}
\newacronym{ehf}{EHF}{extremely high-frequency}
\newacronym{ioe}{IoE}{Internet of Everyone}
\newacronym{gan}{GaN}{Gallium Nitride}
\newacronym{iiot}{IIoT}{Industrial Internet of Things}
\newacronym{dnn}{DNN}{deep neural network}
\newacronym{mae}{MAE}{mean absolute error}
\newacronym{agc}{AGC}{automated guided carrier}
\newacronym{agv}{AGV}{automated guided vehicle}
\newacronym{nwdaf}{NWDAF}{network data analytics function}
\newacronym{rmse}{RMSE}{root mean square error}

\def\BibTeX{{\rm B\kern-.05em{\sc i\kern-.025em b}\kern-.08em
    T\kern-.1667em\lower.7ex\hbox{E}\kern-.125emX}}
    
\begin{document}

\title{Predictive Quality of Service (PQoS): \\ The Next Frontier %of
for Fully Autonomous Systems}
\author{{Mate Boban$^\dagger$,~\IEEEmembership{Member, IEEE}, Marco Giordani$^\dagger$,~\IEEEmembership{Member, IEEE}, Michele Zorzi,~\IEEEmembership{Fellow, IEEE}}

\thanks{$^\dagger$Mate Boban and Marco Giordani are primary co-authors. \newline Mate Boban is with Huawei Technologies Duesseldorf GmbH, Munich Research Center, Germany (email: mate.boban@huawei.com). Marco Giordani and Michele Zorzi are with the Department of Information Engineering (DEI), University of Padova, Italy  (email: \{giordani,zorzi\}@dei.unipd.it).
}}

\maketitle

\begin{abstract}
Recent advances in software, hardware, computing and control have fueled significant progress in the field~of autonomous systems.
Notably, autonomous machines should continuously estimate how the scenario in which they move and operate will evolve within a predefined time frame, and foresee whether or not the network will be able to fulfill the agreed Quality of Service (QoS). If not, appropriate countermeasures should be taken to satisfy the application requirements.
Along these lines, in this paper we present possible methods to enable predictive QoS (PQoS) in autonomous systems, and discuss which use cases will particularly benefit from network prediction. Then, we shed light on the challenges in the field that are still open for future research.
As a case study, we demonstrate whether machine learning can facilitate PQoS in a teleoperated-driving-like use case, as a function of different measurement signals.
\end{abstract}

      \begin{tikzpicture}[remember picture,overlay]
\node[anchor=north,yshift=-10pt] at (current page.north) {\parbox{\dimexpr\textwidth-\fboxsep-\fboxrule\relax}{
\centering\footnotesize This paper has been accepted for publication in IEEE Network, \textcopyright 2021 IEEE.\\
Please cite it as: M. Boban, M. Giordani, M. Zorzi, “Predictive Quality of Service (PQoS): The Next Frontier for Fully Autonomous Systems”, \\ IEEE Network, 2021.}};
\end{tikzpicture}

\section{Introduction}

5G (and beyond) innovations are pushing toward the transition from \emph{automated} to \emph{autonomous} systems to safely operate without human intervention~\cite{shariatmadari2015machine}. For example, driverless cars will be able to mitigate accidents caused by human error, improve the traffic flow, and reduce emissions. 
%Driverless cars are also expected to contribute to a 60\% reduction in emissions, with positive implications for the environment.
On the same wave, collaborative machines in the \gls{iiot} domain~\cite{5gacia2019integration} are making their way into the market to support search and rescue (e.g., military drones), telemedicine (e.g., surgical robots) and workflow optimization (e.g., industrial actuators), as well as to enhance agriculture and logistics (e.g., connected tractors), facilitate diagnostics (e.g., smart home/city sensors), and improve security (e.g., software-defined networks).

Besides novel communication technologies and network architectures, the deployment of autonomous systems is accelerated by recent developments in the areas of \gls{ai} and \gls{ml}, that will make it possible for autonomous devices to gather huge volumes of data and self-optimize~\cite{giordani2020toward}.
However, the dynamic nature of the environment in which the autonomous machines run and operate may significantly complicate network management, which may be unable to fulfill a very demanding set of \gls{qos} requirements (e.g., in terms of very low latency, ubiquitous/robust coverage, and huge downlink/uplink data rate)~\cite{3gppTS22186}.
In particular, unanticipated \gls{qos} degradation may restrict the adoption of safety-critical autonomous applications characterized by extremely high reliability constraints, due to the potentially catastrophic impact of a communication~failure.

In this context, several groups are promoting {\gls{pqos}}, defined as a mechanism to provide autonomous systems with advance notifications about upcoming QoS changes~\cite{5gaa2019qos}. 
%From a network perspective, supporting PQoS implies at least three logical steps: (i) collecting data from different sources; (ii) making QoS predictions based on the acquired data and recognizing upcoming QoS variations; and (iii) informing the intended receivers about network decisions. 
Compared to 5G-like \emph{reactive} strategies, which respond to unanticipated events only after~they occur, a \emph{proactive} behavior gives applications more time to react and permits more efficient network adaptations.
For example, if the upcoming \gls{qos} is not sufficient to handle teleoperated driving, the network may preemptively schedule human takeover to prevent unsafe situations.
Similarly, the resolution of data shared for applications like unsupervised factory control should be dynamically adapted to accommodate bitrate variations, while communication resources should be allocated as a function of the foreseen network capacity.
The time scale of the prediction depends on the use case of interest, as well as the current network dynamics. In general, while for 5G and its predecessors the QoS predictions are typically provided on a slot/frame granularity, future autonomous systems shall support PQoS over a larger time horizon, which may span several minutes or even hours~\cite{5gaa2019qos}. This implies that current prediction mechanisms, based on Bayesian filtering or linear regression~\cite{mason2020adaptive}, should be adapted to permit full-network planning, e.g., leveraging learning-based methods to improve target state estimation~\cite{moreira2020qos}.

\begin{figure*}[!t]
    \centering
    \includegraphics[width=0.99\textwidth]{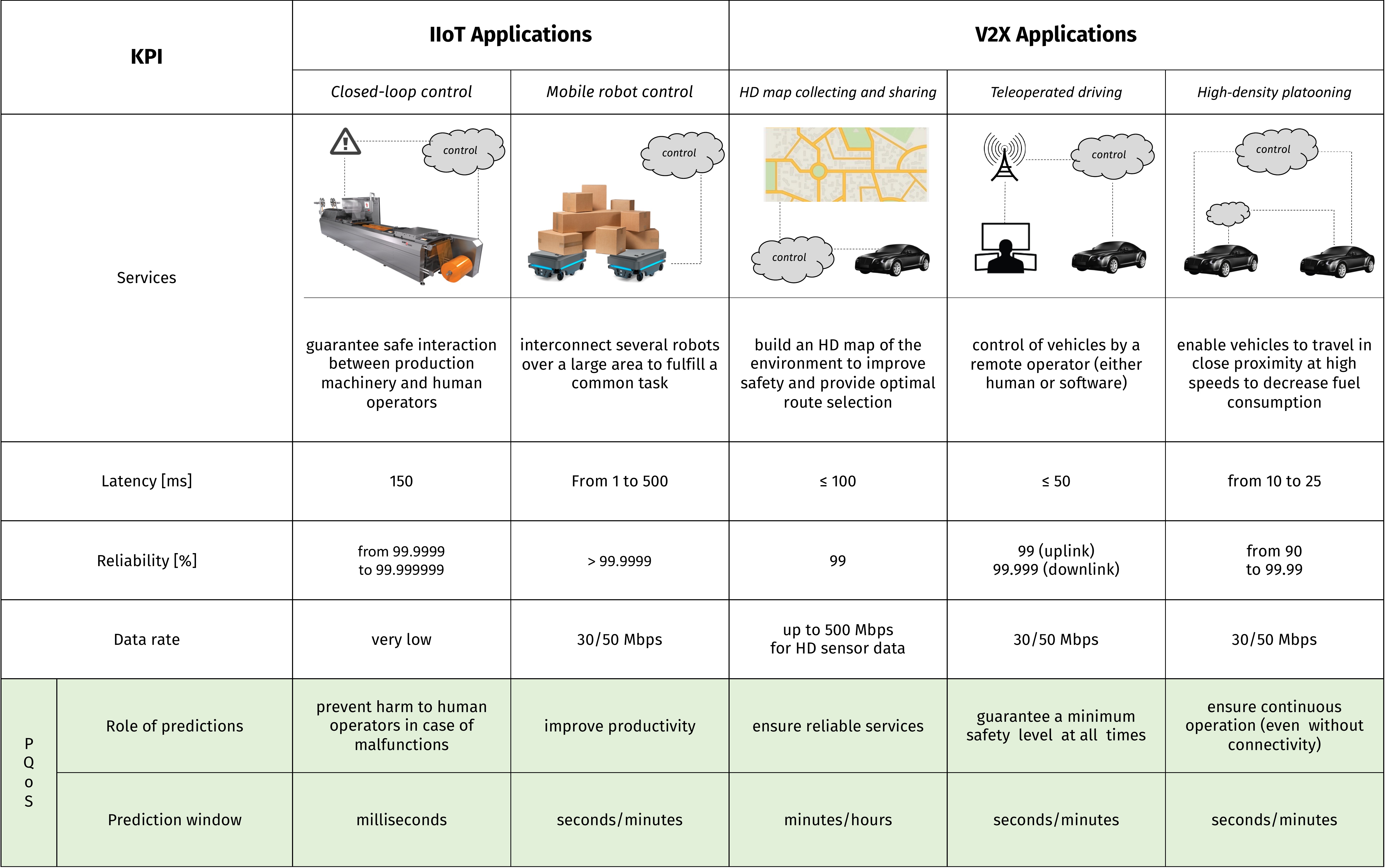}
    \caption{Use cases in the industrial (IIoT)  and vehicular (V2X) domains in which PQoS can be desirable, and relative KPIs according to~\cite{5gacia,22104,5gaa2020cv2x,3gppTS22186}. The last row indicates the prediction window, i.e., the time horizon over which QoS should be predicted and network countermeasures should be taken. }
    \label{fig:use-cases}
\end{figure*}

While \gls{pqos} is generally referred to real-time multimedia/web applications~\cite{cardoso2004quality}, in this paper we characterize PQoS for future autonomous systems.
First, we describe several use cases (and their requirements) for which PQoS is desirable, with a focus on connected autonomous driving and IIoT applications (Sec.~\ref{sec:usecases}). 
Then, we present possible directions to make autonomous systems predictive over a large horizon (Sec.~\ref{sec:tech}). We provide a full-stack perspective with considerations related to both core and radio access network aspects, thus stimulating further research on this topic.
 % also when coordination is unavailable (e.g., during outage situations).
%We shed light on the research challenges associated to PQoS, 
As a case study, we validate the feasibility of predicting QoS (in terms of uplink throughput) of vehicular applications via machine learning (Sec.~\ref{sec:results}). Our results show that \glspl{dnn} can result in more accurate long-term estimates ($80\%$ reduction in average error) compared to a simple linear regression model.
Moreover, we provide numerical evidence of which data should be acquired and processed to facilitate network prediction. We demonstrate that the \gls{sinr} is the most important feature input for PQoS, achieving a prediction error as low as $4\%$ of the maximum throughput range.
%\textcolor{red}{Finally, we identify the future research directions required to harness the potential of PQoS in enabling fully autonomous systems.}

\section{PQoS: Use Cases and Services}
\label{sec:usecases}
%Autonomous systems pose demanding requirements that mandate mobile networks to predict QoS changes in advance and react accordingly.
Originally, system prediction was developed with \emph{functional safety} in mind, to preemptively evaluate the potential safety risks caused by the malfunctioning behavior of electronic/electric systems in vehicles.
Today, autonomous driving systems also implement inter/intra-machine communication, which requires network predictions too, as approved for 5G networks by the \gls{3gpp}~\cite{23288}.  While functional safety expects the system to be shut down in case of system fault, PQoS should support mechanisms to gracefully change the operational mode to ensure that autonomous vertical applications satisfy strict latency/reliability constraints.
In this context, this section reviews some use cases, as illustrated in Fig.~\ref{fig:use-cases}, for which \gls{pqos} is desirable.
Specifically, we focus on (i) the objective of the prediction, (ii) the \gls{qos} Key Performance Indicators (KPIs) to be fulfilled, (iii) the \gls{pqos} window, i.e., the time period for which PQoS analytics is requested, and (iv) the appropriate reactions to be triggered before the predicted \gls{qos} change occurs.

\subsection{PQoS for \gls{iiot} Applications}
\label{ssec:iiot}
Predicting QoS should allow industrial applications to continue operation with adapted functionality under the upcoming QoS conditions. 
In the following, we analyze two industrial automation use cases, selected based on current research trends and the interest in industry.
\smallskip

\paragraph{Closed-loop control} 
Closed-loop control is crucial in industrial processes to guarantee safe interaction between a production machinery and its human operator(s) (e.g., in an assembly line).
Based on environmental sensors, which continuously measure physical parameters (e.g., temperature, pressure) from the production process, the autonomous system configures, monitors, and maintains machines even if the communication service is temporarily unavailable, so as to avoid/minimize accidents and damage of property. % or accidents to human workers.
Due to the critical nature of these applications, determinism becomes a crucial requirement: a $99.99999\%$ reliability is required, while the maximum latency is typically set to around 150 ms, since sensor's measurements are not expected to change rapidly~\cite{5gacia}.

\emph{Role of predictions.} Network predictions are useful to  prevent harm to human operators in case of malfunctions. Based on them, the autonomous systems can shut down quickly enough when reliability is foreseen to change by at least a factor of 10~\cite{5gacia}. For those industrial services that cannot be stopped (e.g., a nuclear power plant), the application may be operated in adapted mode, e.g., at reduced speed or functionality, even though the end-to-end communication service cannot deliver the agreed QoS.
The prediction window should be as short as a few milliseconds to guarantee persistent monitoring of the production station.

%\gls{qos} should be predicted within a short time frame of a few milliseconds, i.e., so that the autonomous system can shut down quickly enough to prevent harm to human operators, while \gls{qos} reports should be broadcast when reliability changes with at least a factor of 10~\cite{5gacia}. 
%For those industrial services that cannot be stopped (e.g., a nuclear power plant), the application may be operated in adapted mode, e.g., at reduced speed or functionality, even though the end-to-end communication service cannot deliver the agreed \gls{qos}.
\smallskip

% Additionally, (safety) control panels are typically equipped with an emergency stop button and an enabling device, which an operator can use in case of a safety event in order to avoid damage. 

%This requires isochronous operation between all robots; in particular, control command messages have to be broadcast amongst the cooperative robots with low jitter. 

\paragraph{Mobile robot control}
An automated production facility may need to interconnect several robots over a large area to fulfill a common task  (e.g., high-precision cooperative carrying~\cite{22104}). 
To do so, mobile robots stream data from their on-board sensors (e.g., cameras and laser scanners) to a remote guidance control system, which then processes the received information and manages the traffic. Sensor data transmission poses severe constraints in terms of throughput (above 10 Mbps~\cite{22104}) which may be difficult to handle with standard communication technologies. Additionally, cooperative machines have to be synchronized in time: the latency requirement ranges from 500 to 1 ms, depending on the degree of~automation~\cite{22104}. 
%Furthermore, the mobility of mobile robots is constrained by their ability to % a maximum flexibility in mobility, which guarantees that they can 
%sense the environment and react timely to the changes. 

%\gls{qos} for mobile robot control should be predicted within a medium/long time frame of a few seconds/minutes, so that robots can reconfigure as the environment changes (e.g., changing the operation speed and/or route) and/or transition to safe mode in a graceful manner. Robots may also adapt their sensors' properties (e.g., the resolution for remote control) to alleviate the burden on the channel during data transmission.
 \emph{Role of predictions.} Network predictions promote improved productivity. Based on them, the system can indeed self-reconfigure (e.g., changing the operation speed and/or route) and/or transition to safe mode in a graceful manner, thus supporting continuous operation even when the communication system is temporarily unavailable. Robots may also adapt their sensors' properties (e.g., the resolution for remote control) to alleviate the burden on the channel during data transmission.
The prediction window is generally consistent with the time frame at which the environment changes, i.e., a few seconds/minutes.

\subsection{\gls{pqos} for \gls{v2x} Applications}
\label{ssec:v2x-use-cases}
 Autonomous driving systems have strict \gls{qos} requirements that need to be satisfied, despite the highly dynamic nature of the \gls{v2x} environment.
In the following, we review three use cases where \gls{pqos} can ensure efficient and reliable driving. 
\smallskip

\paragraph{High-definition map collecting and sharing} 
Autonomous vehicles are equipped with several sensors that gather information from the surrounding environment.
%, including the road topology, vegetation and buildings, and other critical objects, to enhance situational awareness.
More accurate scene understanding can be guaranteed if vehicles broadcast sensors' observations to remote/edge servers. These nodes collect, process, and combine the received data to build a regional \gls{hd} map of the environment to improve safety and provide optimal route selection to the vehicles.
Data sharing, while involving relatively loose latency ($\leq 100$ ms) and reliability ($99\%$) requirements~\cite{5gaa2020cv2x}, may imply multi-gigabits-per-second transmission rates for raw perception data. 

\emph{Role of predictions.} Based on network predictions, the remote server can (i) regulate the number of vehicles required to disseminate HD map updates, so as to reduce network congestion, and (ii) adapt the periodicity of generated messages, to ensure reliable services. The central controller can also adjust the map collection and sharing parameters, such as the level of compression/resolution for sensors' data. A more efficient approach is to select and share only the most valuable environmental elements in the scene (e.g., pedestrians and cyclists), even though this requires the data to be pre-processed before transmission, which may be hard to complete on board of vehicles.
Short-term predictions are desirable, to capture the dynamics of the environment (a prerequisite in safety-critical situations).

%In this scenario, a prediction of the system behavior (e.g., in terms of available time/frequency resources and/or channel conditions) is fundamental to optimize the application parameters and ensure reliable services.
%The remote server can (i) regulate the number of vehicles required to disseminate \gls{hd} map updates, so as to reduce network congestion, and (ii) adapt the periodicity of generated messages. Short inter-transmission intervals are desirable to capture the dynamics of the environment (a prerequisite in safety-critical situations); however, frequent broadcasting may overload the wireless medium and increase packet collisions~\cite{mason2020adaptive}.
%On the other hand, the central controller can adjust the map collection and sharing parameters, such as the level of compression/resolution for sensors' data, which results in a trade-off in terms of transmission efficiency and accuracy. A more efficient approach is to select and share only the most valuable environmental elements in the scene (e.g., pedestrians and cyclists), even though this requires the data to be pre-processed before transmission, which may be hard to complete on board of~vehicles. 
\smallskip

% dire che latency e reliability non sono critical constraints.

% PQoS prediciton lengty

%to extend their perception range beyond the capabilities of their own instrumentation, which involves,

\paragraph{Teleoperated driving} % (fold)
Teleoperated driving enables the control of vehicles by a remote operator (either human or software). The control center must receive/process perception data, including videos from on-board camera sensors, whose data rates can be in the order of hundreds of Mbps for high-resolution transmissions. At the same time, driving decisions from the remote driver must be delivered to the host vehicles with low latency (in the order of 50 ms or less), since higher delays may result in reduced driving responsiveness. 
Moreover, a 99\% (99.999\%) reliability should be supported in uplink (downlink) to receive driving commands~\cite{5gaa2020cv2x}, in addition to other parallel (sensor-based) safety countermeasures to promote more accurate driving operations.

\emph{Role of predictions.} Based on network predictions and in case of QoS degradation, the system should gracefully transition to safe mode, thus guaranteeing a minimum safety level at all times. 
For example, if a certain route is congested, PQoS may assist the control center in adapting the vehicle's speed and trajectory. 
Even though the system should predict the entire end-to-end trip in advance and organize the driving experience accordingly, full-path planning is complicated in dynamic environments, thus making medium-term prediction (e.g., within a time frame of a few minutes) in critical areas (e.g., close to crossroads or intersections) more appropriate.

%In this context, \gls{pqos} allows the application to gracefully degrade to safe mode when needed, and adapt to future conditions, guaranteeing a minimum safety level at all times while still providing optimal performance given the achievable \gls{qos}.   For example, if a certain route is congested, \gls{pqos} may assist the control center in adapting the vehicle's speed and trajectory. %(e.g., in the presence of a congested path, it may be preferable to divert some vehicles towards a different area) 
%Even though the system should predict the entire end-to-end trip in advance and organize the driving experience accordingly, full-path planning is complicated in dynamic environments, thus making medium-term prediction (e.g., within a time frame of a few minutes) in critical areas (e.g., close to crossroads or intersections) more appropriate.
\smallskip

\paragraph{High-density platooning} 
 Platooning enables vehicles to cooperatively travel in close proximity at high speeds, to decrease fuel consumption and traffic congestion~\cite{3gppTS22186}.
 Platooning can be supported by a cloud assistant application, which gathers status information from the platoon head/leader, such as common mobility patterns (e.g., speed, heading, intended maneuvers) and perception data from on-board sensors, and decides the behavior and configuration of the platoon.
 \gls{qos} requirements are given according to the desired degree of automation: while the required latency for collecting and processing status updates and broadcasting platooning decisions ranges from 10 to 25 ms depending on the distance, the required reliability ranges from $90\%$ to $99.99\%$~\cite{3gppTS22186}.
 
 \emph{Role of predictions.} Network predictions are useful to estimate when platoon members are out-of-coverage, and ensure continuous operation even without connectivity. Based on them, the cloud assistant can manage the platoon formation, e.g., regulating join/leave operations, the number of platoon members, the distance between vehicles and their speed. It can also control local perception parameters, such as reducing the number of vehicles required to share sensors' data to mitigate the traffic load on the channel. In case QoS degradation or outage (e.g., in tunnels) is foreseen, the cloud assistant may instruct the platoon leader to support driving activities by coordinating resources autonomously over the sidelink, or request human intervention.
A medium/long prediction window of a few seconds/minutes is beneficial to efficiently capture the road dynamics.
 
 %\gls{qos}  should be estimated over a time horizon of seconds or minutes, to efficiently capture the road dynamics and maintain safety.
%After predictions, the cloud assistant may undertake several countermeasures to manage the platoon formation, e.g., regulating join/leave operations, the number of platoon members, the distance between vehicles and their speed. It can also control local perception parameters, such as reducing the number of vehicles required to share sensors' data to mitigate the traffic load on the channel. In case \gls{qos} degradation or outage (e.g., in tunnels) is foreseen, the cloud assistant may instruct the platoon leader to support driving activities by coordinating resources autonomously over the sidelink (referred to as \emph{resource allocation mode 2} in the 3GPP terminology for NR V2X), %~\cite{zugno2020toward}) 
%or request human intervention, so as to continue the operation even without cellular~connectivity.

\section{Enabling PQoS: Technologies and Trends}
\label{sec:tech}
Besides introducing PQoS procedures in Sec.~\ref{ssec:alg}, we distinguish between \gls{cn} and \gls{ran} aspects in Sec.~\ref{sec:PQoSCore} and \ref{ssec:ran}, i.e., the two main 4G/5G system architecture components.

\subsection{Algorithms and Procedures}
\label{ssec:alg}
%Three algorithmic approaches presented in WiLAB slides: for short, medium, long term prediction. Would ideally add a figure here...

\begin{figure}[!t]
    \centering
    \includegraphics[width=\columnwidth]{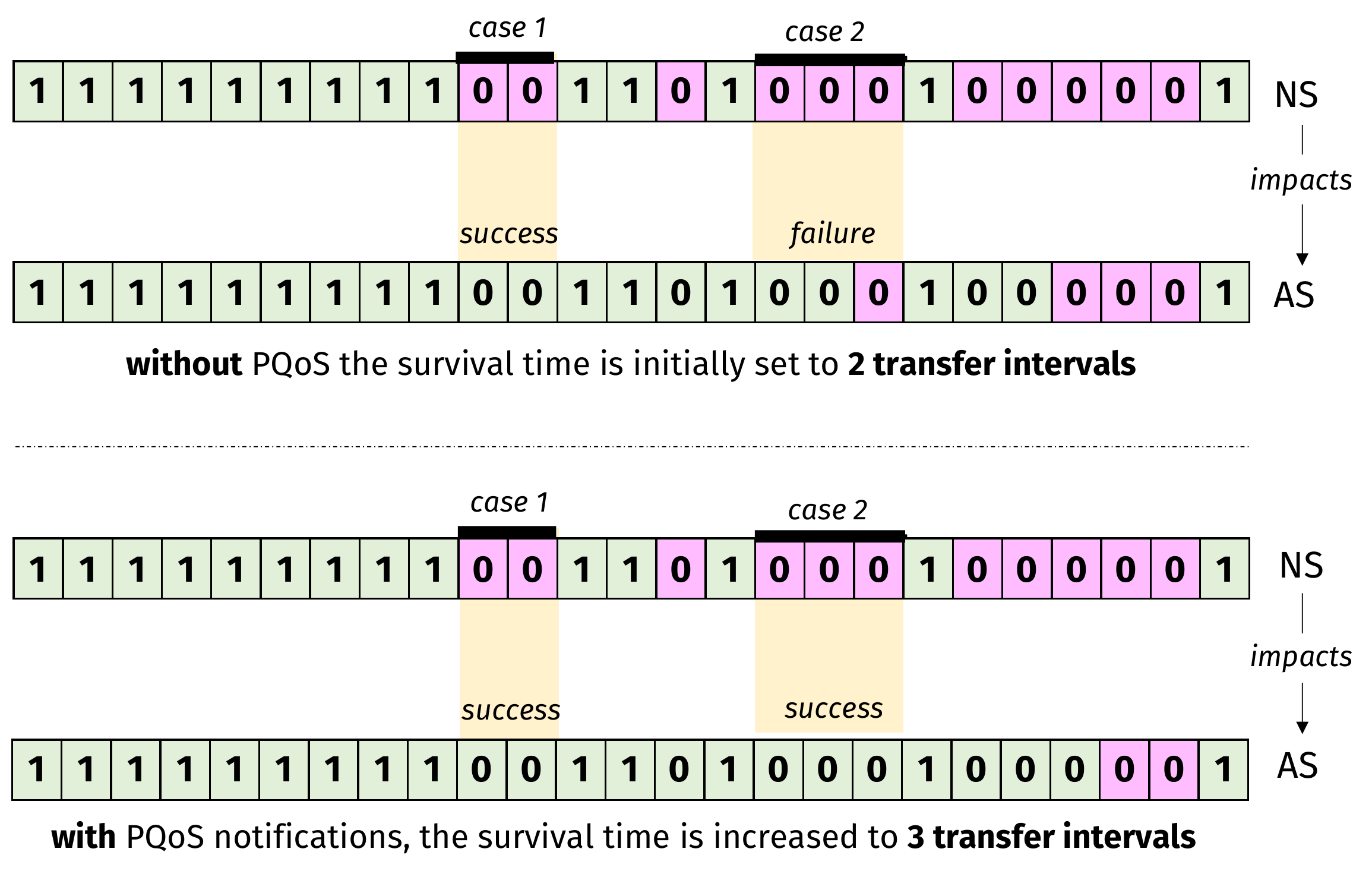}
    \caption{Comparison of network service (NS) and application service (AS) failures with and without PQoS. If notifications about future network service failures are exchanged, the system can increase the survival time to three transfer intervals, thus ensuring no disruption.}
    \label{fig:PQoSShort}
\end{figure}

According to the 5G-ACIA, dependability in autonomous systems is crucial to provide information about the network changes \textit{ahead of time}.
In this context, PQoS enables the autonomous system to take actions \textit{before} machines go into a down/failure state. In this condition, either the data units do not arrive correctly within a given latency constraint (i.e., network service failure), or the application cannot continue to function, given its settings and the network status (i.e., application service failure). We define the \emph{survival time} as the time that the application service can ``survive'' without the network service. 
As an example, in the top part of Fig.~\ref{fig:PQoSShort} (where each ``0'' or ``1'' represents a failure or success for one transfer interval, respectively) the application service has a survival time of two transfer intervals and does not leverage advance notifications of network service. In this scenario, the application service can ``survive'' two consecutive failures (depicted as ``case 1'' in the figure), but cannot survive three (depicted as ``case 2''). In the bottom part of Fig.~\ref{fig:PQoSShort}, we now assume that \gls{pqos} provides advance notifications several transfer intervals ahead of time. %(e.g., in case of an IIoT application, by using the trajectory of a robot to predict an upcoming deep fade on the wireless link).
This would allow the application service to adapt to ``case 2'' by regulating its survival time to three transfer intervals, thereby ensuring that there is no disruption, at the expense of reduced \gls{qos} performance.  

From the example above, it appears clear that it is vital to correctly estimate the time evolution of the network and react accordingly. The optimal strategy depends on the time horizon of the prediction, as well as on its ``quality'' (e.g., in the form of probabilistic confidence interval of the predicted value).
While for short-term prediction (e.g., in the context of functional safety) faults occur mainly due to hardware imperfections, in case of longer-term predictions targeted by %for which software analysis is preferred, for 
PQoS \gls{ml}, radio and network conditions are the main culprit for faults: PQoS aims at providing methods able to predict the network behavior with high accuracy. Compared to traditional prediction algorithms like linear regression and filter-based models, which do not efficiently handle large amounts of input data or are unable to provide predictions for non-linear relationships between features and labels, ML solutions hold the promise of %like \glspl{dnn} provide
more accurate predictions~\cite{yin2020qos}. % outcomes thanks to hidden layers.
Notably, DNNs are able to learn meaningful relationships based on network observations (including radio and network conditions, available resources, predicted mobility patterns, etc.), and return the possible actions to sustain \gls{qos} in case of unseen conditions without pre-programmed/a-priori rules. Furthermore, methods such as transfer learning~(TL) can be leveraged to transfer the knowledge acquired from one scenario to another 
(e.g., in which communicating nodes are deployed in different locations), without the need to learn from scratch.

However, applying ML-based approaches introduces several trade-offs. On one side, model training requires significant computational power and time, together with the availability of a large amount of data to build robust predictors.  Additionally, model retraining, due to data distributions drifting over time in dynamic environments, involves additional overhead, which may not be tolerated by safety-critical use cases.
Nevertheless, the above issues can be mitigated by adopting offline (rather than online) training, %e.g., (un)supervised learning, 
or by delegating the training operation to external processors: once the model is trained, inference can be done even on devices %on board of the autonomous system, 
with limited computational requirements.

\subsection{\gls{pqos} Core Network (CN) Aspects}\label{sec:PQoSCore}
In terms of the \gls{cn} (responsible for end-to-end connection management and Internet access), the initial work on \gls{pqos} has been done by vertical industry associations. Specifically, following the initial pre-standardization efforts by 5GAA~\cite{5gaa2019qos} and 5G-ACIA~\cite{5gacia}, the 3GPP defined the so-called \gls{nwdaf} to provide analytics services to 5G \gls{cn} functions~\cite{3gppTS29520}. \gls{nwdaf} is  a key enabler for network automation, aggregating data, and providing analytics using either a request or a subscription model.  \gls{nwdaf} enables %at least 
several functionalities~\cite{3gppTS23791} that are \gls{pqos}-related. They can be grouped as follows:
\begin{itemize}
    \item \emph{\gls{qos} provisioning and adjustment}, designed to support the requirements that different \gls{qos} parameters and traffic types mandate from the network, along with their timely adaptation and related data traffic handling;
    \item  \emph{mobility/topology-related} management, to assist edge computing functions, and identify the area(s) of network topology that might have oscillating \gls{qos}; 
    \item  \emph{policy adjustment}, designed to support sharing of analytics, determine the appropriate policy based on them, and ultimately ensure predictable network performance.
\end{itemize}

While not yet fully developed in 3GPP specifications, the above functionalities are a good starting point to enable \gls{pqos}. 
%For example, by tightly integrating the \gls{nwdaf}  \gls{qos} provisioning and adjustment services, the \gls{pqos} notification can be efficiently generated upon request. 
Whether \gls{nwdaf} should be implemented in software or hardware is still unclear. Although the former approach permits to host \gls{nwdaf} facilities at low cost and closer to the end machines, %(e.g., on the cloud), 
thus promoting lower latency, the latter has access to more computational resources, at the expense of~flexibility.

\subsection{\gls{pqos} \gls{ran} Aspects}
\label{ssec:ran}
Having \gls{pqos} enabled in the \gls{cn} only will not suffice for safety-critical and time-critical applications that are frequent %as those 
in \gls{iiot} and \gls{v2x}. To reduce the latency and increase the flexibility, %In these regards, prediction shall be also implemented in the \gls{ran}. 
%To date, there has been scarce work on \gls{pqos} in \gls{ran}. 
following the architectural aspects and \gls{cn} functions defined in the scope of \gls{nwdaf}, standardization activities have started in \gls{ran} Working Group 3, where 
%(dealing with architectural aspects of \gls{ran}) 
a study item has recently been approved to define the high level principles for \gls{ran} intelligence (mainly responsible for radio-resource handling and data transmission) enabled by \gls{ai}~\cite{3gppRP201304}. 
%This process will develop a functional framework and identify the benefits of \gls{ai} on the \gls{ran} side. 
%The use cases it initially targets are energy saving, load balancing, mobility management, and coverage optimization. 
While %the research is 
still in its infancy, the study promises to support \gls{pqos} functionalities.
%at least for the short- and medium-term~predictions.

In terms of technical enablers in RAN-side prediction, we highlight the following promising approaches.
\begin{itemize}
    \item  \emph{Radio map prediction}.
    Predicting the physical aspects of radio propagation (e.g., path loss or \gls{sinr}, that is generating ``radio maps'') shows promising results~\cite{levie2019radiounet}. Specifically, using measurements or ray tracing as ``ground truth'' for radio maps and applying \gls{ml}-based approaches to predict the maps in foreseen locations can serve as an underlying ``layer'' for throughput, latency, and other application-level \gls{qos} parameters.
    \item  \emph{Predictive adaptation of radio parameters}. The \gls{ran} should convert QoS estimates into appropriate network decisions if service requirements are not satisfied. Besides operating directly on the mobility %driving 
    patterns, the \gls{ran} may undertake lower-layer actions, e.g., changing communication mode, adapting the radio resource allocation based on the available network capacity, or modifying the system numerology and/or the communication spectrum.
    \item  \emph{Predictive scheduling}. The \gls{ran} should make 
    %~\cite{sahin2018reinforcement}, 
    scheduling decisions in advance of the network changes. Depending on whether or not network coverage will be guaranteed in future time instants and geographical locations,  resources can be scheduled either by the base station or autonomously by the machines, respectively.
\end{itemize}

%The application \gls{qos} is the final result of the functioning of all the network layers, which makes predicting \gls{qos} a complex topic. 
In terms of \gls{ran}-based prediction, while lower-layer metrics, such as the users' locations or the \gls{sinr}, may be able to capture both signal strength and propagation conditions, high-layer parameters (e.g., average data rate, end-to-end latency, or packet delivery ratio) may be used to train more accurate prediction models. In Sec.~\ref{sec:results} we evaluate which data can best support \gls{qos} predictions for a specific scenario. 

\section{PQoS Performance Evaluation: A Case Study}
\label{sec:results}
In Sec.~\ref{sec:tech} we reviewed how machine learning could facilitate network prediction. As a case study, we now evaluate the ability of machine learning to predict the QoS of V2X applications using measurement data.
We describe our measurement scenario in Sec.~\ref{sub:field_trial_and_parameters}, and present the numerical results in Sec.~\ref{sub:numerical_results}.
\subsection{Measurement Scenario and Parameters} % (fold)
\label{sub:field_trial_and_parameters}

\begin{figure}[!t]
    \centering
    \includegraphics[width=\columnwidth]{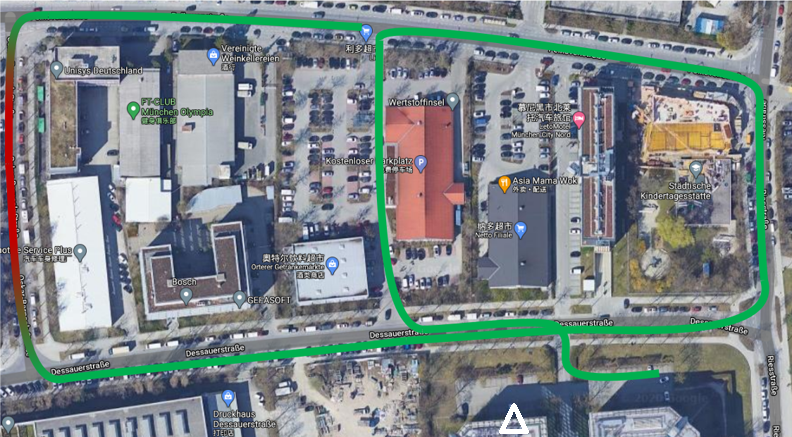}
    \caption{Map of the location where the measurements were collected. The route followed by the test vehicle is marked with an overlay plot, showing warmer color (red) for lower throughput and colder color (green) for higher throughput. The base station antenna was located 21 meters above the ground level, on top of the building marked with the white triangle, whereas the vehicle antenna was at a height of approximately 1.5 meters. The test vehicle traveled the loop shown in the figure 10 times.}
    \label{scenario}
\end{figure}

Our measurement campaign was performed in Munich, Germany. The scenario (illustrated in Fig.~\ref{scenario}) was defined so as to resemble the teleoperated driving use case (Sec.~\ref{ssec:v2x-use-cases}), in which vehicles transmit video stream(s) to the teleoperating center with a maximum uplink throughput of 40 Mbps~\cite{5gaa2020cv2x}. The measurements were collected at 3.41 GHz, with 40 MHz of bandwidth and antenna gain of 15.5 dBi (5 dBi) at the base station's (user's) side. We collected the following information, referred to as ``input features'' in the remainder of this paper: uplink/downlink throughput, \gls{sinr} (measured as the ratio of the signal power to undesired interference plus noise power) and \gls{rsrp} (an indication of the channel quality at the carrier level) for each base station antenna sector, location (longitude, latitude, altitude) and velocity of the vehicle.
The measurements were collected on a relatively small loop with stable traffic/interference conditions. This makes our input features able to predict the system for any time horizon. In a more dynamic environment, the applicable time horizon will depend on the rate of variability of these parameters.

The collected dataset was used to predict the throughput using several machine learning approaches. In particular, we employed linear regression and \glspl{dnn}, and used different input features to evaluate which of them (or their combination) resulted in the best prediction of the uplink throughput.
The motivation for using these two methods is to evaluate whether general-purpose prediction methods -- one more computationally simple (linear regression) and the other more complex but potentially broadly applicable (\gls{dnn}) -- can be effective at predicting the uplink throughput.

The \gls{dnn} we used for prediction consists of a normalization layer, followed by three dense layers. The first two dense layers have 64 outputs and use ``ReLu'' activation, with the final dense layer resulting in one output (i.e., the predicted throughput value). The collected measurement dataset, containing approximately 3100 samples, was divided into training and testing datasets. The training dataset contained 90\% of the randomly selected samples, with the remaining 10\% used for testing.

\subsection{Simulation Results}
\label{sub:numerical_results}
We first explored the relative benefit of each of the input features. SINR and location emerged as the dominant features in case of both linear regression and DNN. This is expected since (i) the SINR is proportional to the throughput (with SINR above 10 dB nearly always resulting in the maximum achievable throughput of 40 Mbps),
and (ii) both training and testing datasets were collected along the same route, and thus incorporate the same location information. Furthermore, the remaining input features have been observed to be insignificant or even detrimental in the presence of SINR and distance features, due to reduced learning rate. Specifically, the speed of the user was not high enough to exhibit measurable variations due to Doppler, whereas the RSRP per sector was better accounted for as part of the SINR. 
Moreover, despite the separation in frequency, the downlink SINR and throughput have shown high correlation with their respective uplink counterparts, indicating that the dominating factor on the channel was large scale fading. 
%This also made the downlink SINR and throughput redundant, since uplink SINR and throughput contained virtually the same information. 
%We verified this by testing with and without downlink SINR and throughput as input features, which resulted in nearly identical predictions.

Based on these early results, we explored six prediction models with the following combinations of input features:
\begin{itemize}
\item Linear regression (or DNN) with location only;
\item Linear regression (or DNN) with SINR only;
\item Linear regression (or DNN) with both SINR and location.
\end{itemize}
%but they showed no significant benefit; I think this will be more clean and will not complicating things in terms of detailed feature descriptions
%This leaves us with: 
%$linear_SINR_model$                  
%$linear_location_model$            
%$linear_model             $               
%$dnn_location_model     $          
%$dnn_SINR_model          $           
%$dnn_model                    $           
%I played with different way to plot the results, and decided to plot all error curves on one figure. See attached for an example. Furthermore, I think we can use just one Prediction vs measurement plot (e.g., for best, DNN model) or maybe two (DNN full, and DNN SINR). This could potentially save one figure/table for other purpose (e.g., including the below testing performance results as a table).

\begin{table}[t!]
\centering
\caption{Mean absolute and relative error on {testing} data for different prediction~methods.}
\label{MAETable}
\renewcommand{\arraystretch}{1.4}
\begin{tabular}{ m{4.2cm} M{1.4cm} M{2cm}}
%\begin{tabular}{lll}
 %\hline
% \multicolumn{2}{|c|}{Mean absolute error (Mbps)} \\
 \hline
\textbf{Prediction algorithm %Mean absolute error (MAE)
}&\textbf{MAE (Mbps)} &\textbf{Avg. throughput error (\%)}\\
 \hline
Linear regression, location only &                                     2.21 & 6.5~\% \\
Linear regression, SINR only &                                                    2.19 &6.4~\%\\
Linear regression, SINR and location           &                                    2.08&6.1~\%\\
DNN, location only                &                                          0.92&2.7~\%\\
DNN, SINR only &                                              0.46&1.4~\%\\
DNN, SINR and location           &                                        0.40&1.2~\%\\
 \hline
\end{tabular}
\end{table}

\begin{figure}[!t]
    \centering
    \includegraphics[width=0.9\columnwidth]{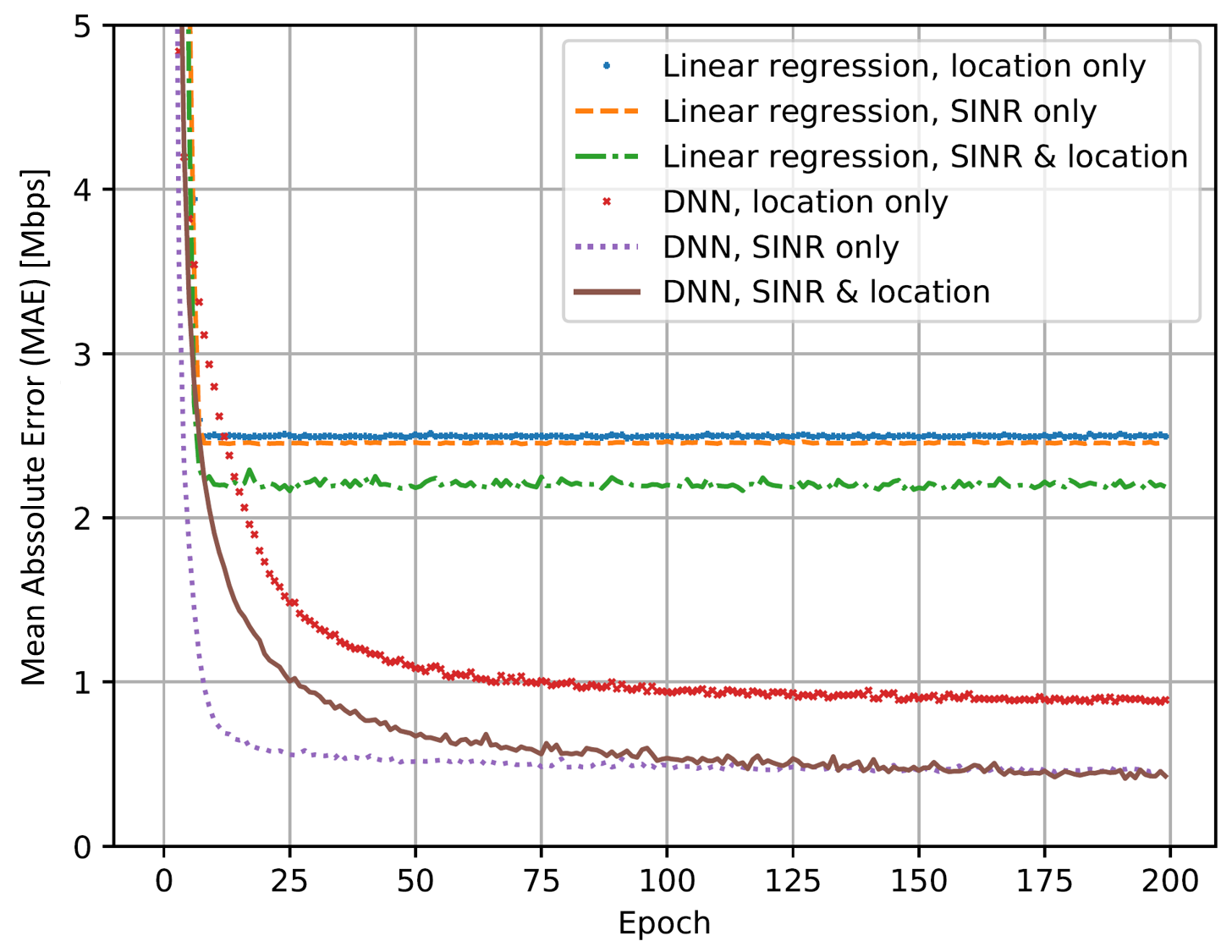}
    \caption{Training error (in terms of MAE) for different prediction approaches of the uplink throughput, as a function of training epoch, i.e., the number of cycles through the full training dataset.}
    \label{trainingError}
\end{figure}

Fig.~\ref{trainingError} shows the training error, measured in terms of \gls{mae}, of the uplink throughput over the duration of training for the six models. All linear models converge very quickly to their best performance (within 20 epochs), irrespective of which features are used. 
Notably, the combination of SINR and location results in a markedly better performance at virtually no expense of further training. However, all linear regression models converge to a minimum of about 2 Mbps MAE (as reported in Table~\ref{MAETable}). This indicates that the linear model is overly simple in describing the relationship of the input features and the predicted uplink throughput, thus calling for more advanced approaches, as discussed in Sec.~\ref{ssec:alg}. 
Specifically, linear regression optimizes predictions only for throughput samples above 35 Mbps, which make up almost 90\% of the data we collected.

\begin{figure}[!t]
    \centering
    \includegraphics[width=\columnwidth]{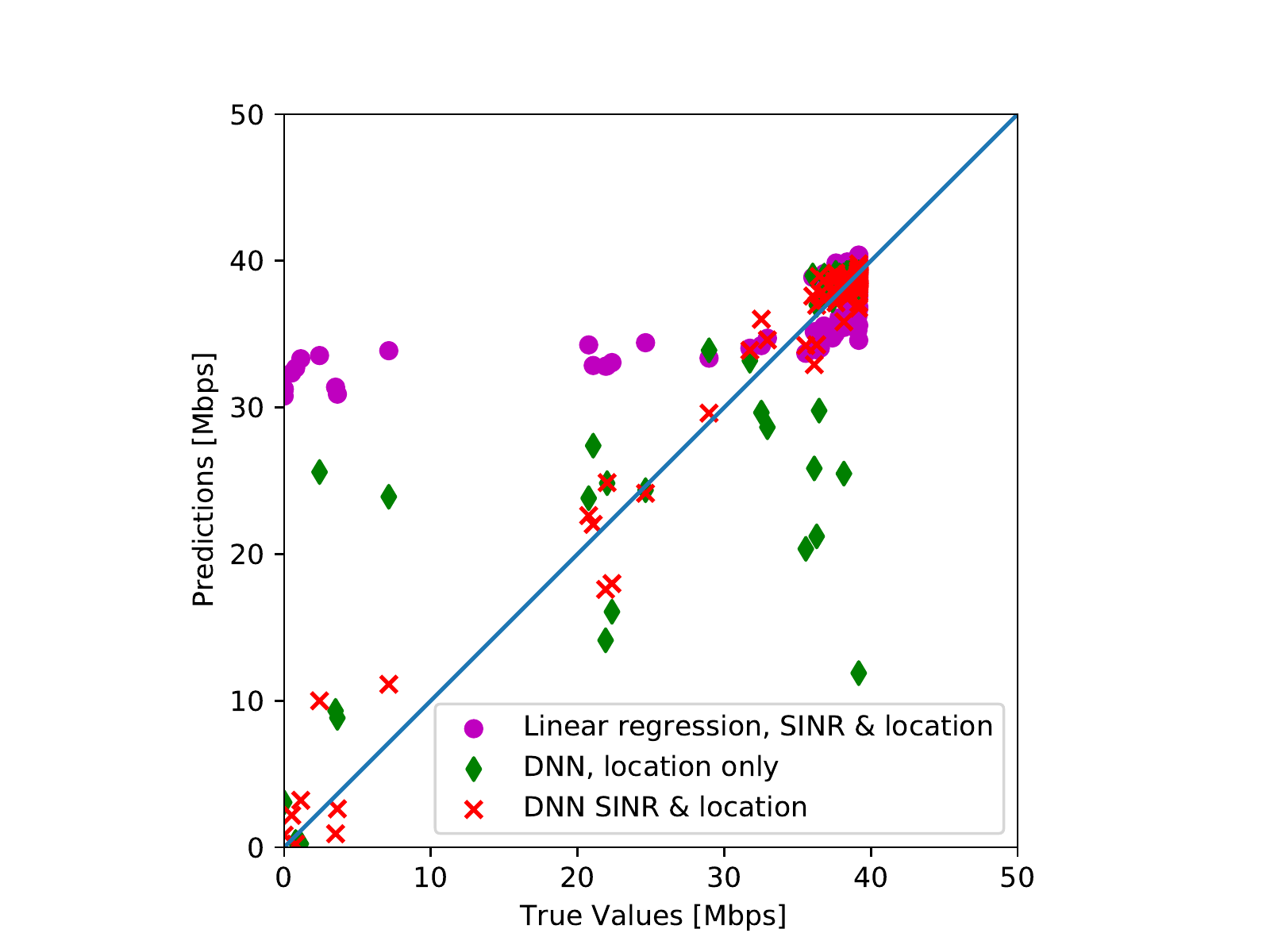}
    \caption{Predictions vs true values (measurements) in the testing uplink throughput dataset. For readability, we omit 
the results for "Linear regression, location only" and "Linear regression, SINR only" (both similar to "Linear regression, SINR and location," with slightly higher error). We also omit "DNN, SINR only" (similar to "DNN, SINR and location") results.}
    \label{fig:prediction}
\end{figure}

On the other hand, while all three DNN models require increased training (over 50 epochs), they converge to below 1 Mbps MAE. 
In particular, the ``DNN, location only'' model results in MAE of 0.92 Mbps, while the ``DNN, SINR only'' model further reduces the MAE to 0.46 Mbps (see Table~\ref{MAETable}) with a considerably faster rate of convergence. This indicates a more direct relationship between SINR and throughput.
%and a smaller size of the feature space (each SINR sample is a single entry, whereas location is composed of three entries, namely longitude, latitude, and altitude).
Finally, the ``DNN, SINR and location'' model ultimately results in the lowest MAE of 0.4 Mbps, even though this requires 150 epochs of training compared to around 50 for the ``DNN, SINR only'' model.
While this does not affect the real-time prediction performance, since the training phase is performed offline, convergence time should still be minimized as much as possible to facilitate computationally efficient predictions and promote faster retraining.
%Its rate of convergence is more akin to the ``DNN, location only'' case, indicating that adding location information, while ultimately resulting in a lower error, comes at the expense of increased training. Therefore, a trade off exists: while ``DNN, SINR only''  converges considerably faster (see Fig.~\ref{trainingError}), it is eventually (and with sufficient training) further assisted by location features  (see Fig.~\ref{trainingError} and Table~\ref{MAETable}). 

Similar conclusions can be drawn from Fig.~\ref{fig:prediction}, which shows the comparison of the measured values of throughput in the testing dataset and the prediction by different algorithms. Due to the smaller number of samples in the lower range of throughput, all linear regression approaches result in high error. On the other hand, DNN approaches result in both lower and more uniform error across the range of throughput values.
For completeness, Table~\ref{MAETable} also shows the average throughput error, calculated as the ratio between the \gls{mae} and the mean  uplink throughput of the collected dataset (34~Mbps). 
%\textcolor{red}{We also evaluated the performance of piece-wise linear regression. While this approach resulted in a lower error, it came at a cost of increased training time comparable to that of DNN, while still yielding MAE higher than that of DNN.}

Our results indicate that SINR is the single most important feature, among those collected, when it comes to predicting throughput. Notably, SINR can better generalize to unseen locations, as it incorporates in a single value both the properties of a certain location and the transmitter/receiver distance. Moreover, %unlike location information, 
SINR %is time-invariant, and 
subsumes any temporal effects of the channel, which helps with long-term predictions, a fundamental prerequisite for PQoS applications.
%In other words, the location information will be useless if throughput needs to be predicted in unseen geographical locations. 
%
%This is why building precise radio maps of an area, as currently under investigation within 3GPP RAN activities (see Sec.~\ref{ssec:ran}), is of high importance in predicting the throughput. 
We further explored the impact of reducing the precision of the collected SINR (i.e., the accuracy of radio maps) 
by introducing noise.
Specifically, increasing an SINR \gls{rmse} from 0 to 4 still resulted in a MAE of less than 1 Mbps, whereas an SINR RMSE of 10 resulted in 1.5 Mbps MAE. This is a positive result, since it indicates that radio maps do not need to be perfect to still obtain a reasonably good throughput prediction (i.e., within 4\% of the maximum throughput range).

\section{Conclusions and Next Steps}
Based on the current research and industry alliance efforts, we analyzed the ability of \gls{pqos} to accurately predict the network performance of autonomous systems, as a means to gracefully change their operational mode if the target performance cannot be achieved. We discussed the techniques -- most notably those relying on machine learning -- that help enable \gls{pqos} in the core and radio access parts of the network. %with the prediction windows ranging from short, to medium, to very long. 
Notably, we draw the community's attention to the development of mechanisms to adjust end users' mobility/topology patterns, scheduling decisions, and radio parameters, to satisfy service requirements in view of the predicted QoS. We exemplify the potential benefits of \gls{pqos} through a real-world example of predicting uplink throughput in \gls{v2x} via \gls{ml} techniques. Our simulations prove that DNNs based on SINR measurements can result in accurate predictions, while involving longer training than a linear regression approach.

As identified in Sec.~\ref{sec:tech}, there is a lot of work remaining for \gls{pqos} to fulfill its promise. In particular, techniques that allow \gls{ran}-based predictions need to be enabled, in particular those that can provide benefits using relatively small and local input information. Furthermore, a tighter integration between \gls{ran} and core network is required, with prediction information flowing both ways for some use cases (e.g., teleoperated driving). Finally, little work has been done on incorporating the sidelink into the \gls{pqos} ecosystem. Combined with distributed \gls{pqos} algorithms, sidelink holds the promise of low latency, local prediction with minimum uplink/downlink control from the network.

\bibliographystyle{IEEEtran}

\begin{IEEEbiographynophoto}{Mate Boban}
[SM'21] received the Ph.D. degree in electrical and computer engineering from Carnegie Mellon University, Pittsburgh, PA, USA. He is with Huawei Technologies, Munich Research Center, Germany. He is currently an Associate Editor of IEEE Transactions on Mobile Computing. He coauthored three papers that received the Best Paper Award. His current research interests include resource allocation, machine learning applied to wireless communications, and channel modeling.
\end{IEEEbiographynophoto}

\begin{IEEEbiographynophoto}{Marco Giordani}
[M'20] received his Ph.D. in Information Engineering in 2020 from the University of Padova, Italy,  where he is now a postdoctoral researcher and adjunct professor.
He visited  NYU and TOYOTA Infotechnology Center, Inc., USA.
In 2018 he received the “Daniel E. Noble Fellowship Award” from the IEEE Vehicular Technology Society. His research  focuses on protocol design for 5G/6G mmWave cellular and vehicular networks.
\end{IEEEbiographynophoto}%

\begin{IEEEbiographynophoto}{Michele Zorzi}
[F'07] is with the Information Engineering Department of the University of Padova, focusing on wireless communications research. He was Editor-in-Chief of IEEE Wireless Communications from 2003 to 2005, IEEE Transactions on Communications from 2008 to 2011, and IEEE Transactions on Cognitive Communications and Networking from 2014 to 2018. He served ComSoc as a Member-at-Large of the Board of Governors from 2009 to 2011 and from 2021 to 2023, as Director of Education and Training from 2014 to 2015, and as Director of Journals from 2020 to 2021.
\end{IEEEbiographynophoto}

\end{document}